\documentclass[12pt,preprint]{aastex}

\usepackage{datetime}

\usepackage{amsmath}
\usepackage{verbatim}
\usepackage{wasysym}
\usepackage{multirow}
\usepackage{graphicx}
\usepackage{soul} 
\usepackage{color}

\definecolor{lightgray}{rgb}{.9,.9,.9}

\usepackage[T1]{fontenc}
\usepackage{times}

\begin{document}

\title{Delivery of dust grains from comet C/2013~A1 (Siding Spring) to Mars}

\author{Pasquale Tricarico, Nalin H.~Samarasinha, Mark V.~Sykes, Jian-Yang Li}
\affil{Planetary Science Institute, \\ 1700 E.~Ft.~Lowell \#106, Tucson, AZ 85719, USA}
\email{\href{mailto:tricaric@psi.edu}{tricaric@psi.edu}}

\author{Tony L.~Farnham, Michael S.P.~Kelley}
\affil{Department of Astronomy, University of Maryland, \\ College Park, MD 20742, USA}

\author{Davide Farnocchia, Rachel Stevenson, James M.~Bauer, Robert E.~Lock}
\affil{Jet Propulsion Laboratory, California Institute of Technology, \\ 4800 Oak Grove Drive, Pasadena, CA 91109, USA}

\begin{abstract}
Comet C/2013 A1 (Siding Spring) will have a close encounter with Mars on October 19, 2014.
We model the dynamical evolution of dust grains 
from the time of their ejection from the comet nucleus to the Mars close encounter,
and determine the flux at Mars.
Constraints on the ejection velocity from Hubble Space Telescope observations indicate that the 
bulk of the grains will likely miss Mars, although it is possible that a few-percent of grains with higher velocities will reach Mars, peaking approximately 90--100 minutes after the close approach of the nucleus,
and consisting mostly of millimeter-radius grains ejected from the comet nucleus
at a heliocentric distance of approximately 9~AU or larger.
At higher velocities, younger grains from sub-millimeter to several millimeter can reach Mars too,
although an even smaller fraction of grains is expected have these velocities,
with negligible effect on the peak timing.
Using NEOWISE observations of the comet,
we can estimate that the maximum fluence will be of the order of $10^{-7}$ grains/m$^2$.
We include a detailed analysis of how the expected fluence
depends on the grain density, ejection velocity, and size-frequency distribution,
to account for current model uncertainties and in preparation of possible refined model values in the near future.
\end{abstract}

\keywords{comets: individual (C/2013 A1 (Siding Spring)) --- meteorites, meteors, meteoroids}

\maketitle

\section{Introduction}

Comet C/2013 A1 (Siding Spring) will have a close approach (hereafter c/a) 
with Mars on October 19, 2014 at approximately 18:29 UT,
reaching a minimum distance of approximately 134,000~km from the center of Mars,
according to JPL solution \#46 \citep{Farnocchia_in_prep}.
This orbital solution was retrieved from the JPL Horizons system \citep{1996DPS....28.2504G}, on March 17, 2014, and include all the observations before the low solar elongation period.
The comet is on a hyperbolic retrograde orbit (129$^\circ$ inclination), and the encounter with Mars will be at 
a relative velocity of approximately 56 km/s.
Observed from Mars, the comet reaches a minimum solar elongation of approximately 72$^\circ$.
After the close approach between the two bodies, Mars continues to move closer to the orbit of the comet,
crossing the comet orbit plane approximately 100 minutes after c/a and then reaching
a minimum distance of approximately 27,000~km from the comet orbit approximately 102 minutes after c/a.
A detailed trajectory analysis by \cite{Farnocchia_in_prep}
indicates that the current uncertainty on the c/a distance, 
projected on the target plane normal to the velocity of the comet nucleus relative to Mars,
is an ellipse approximately $1600 \times 700$~km (1$\sigma$)
with the long axis 20$^\circ$ from the line connecting the comet nucleus and Mars,
and the c/a time uncertainty is about 1 minute (1$\sigma$).
These uncertainties exclude any collision of the comet nucleus with Mars,
yet warrant a detailed investigation of the possibility
that the dust coma of the comet may be able to reach Mars.

\begin{figure}[t]
\begin{center}
\includegraphics*[width=0.7\textwidth]{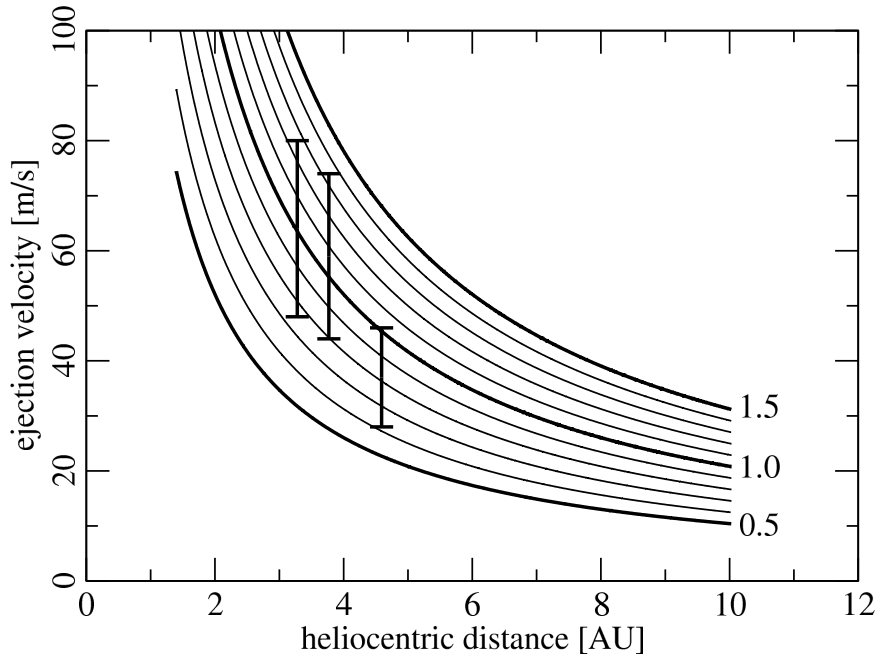}
\end{center}
\caption{Ejection velocity of dust grains with $\beta=1$ as a function of the heliocentric distance
of the nucleus at the time of the ejection.
The curves are not a fit, but correspond to the empirical scaling law of Eq.~\eqref{eq:v_ej} 
with $v_\text{ref}$ values between 0.5~m/s and 1.5~m/s in increments of 0.1~m/s.
The three vertical bars are the velocity ranges as constrained from HST observations \citep{Li_in_prep}.
}
\label{fig:v_ej_micron_vs_Rh}
\end{figure}

C/2013~A1 was discovered at a heliocentric distance of 7.2~AU \citep{2013CBET.3368....1M},
suggesting hyper-volatile activity (i.e., driven by CO or CO$_2$)
was likely causing long-term build-up of larger grains in the coma. 
Hubble Space Telescope (HST) observations by \cite{Li_in_prep} allow us to estimate the ejection velocity of the dust grains at the $\mu$m radius scale,
or more precisely grains with $\beta\simeq1$,
where $\beta$ is the ratio between the radiation pressure force and the gravitational force due to the Sun on the grain \citep{1979Icar...40....1B}.
Observations were performed on October 29, 2013, on Jan 21, 2014, and on March 11, 2014.
By using a technique based on the sunward turn back distance in the continuum \citep[e.g.,][]{2013Icar..222..799M} 
we determined 
a velocity approximately between 28 and 46~m/s at a heliocentric distance of 4.59~AU, 
then between 44 and 74~m/s at 3.77~AU,
and between 48 and 80~m/s at 3.28~AU,
all for $\beta=1$ dust grains, see Figure~\ref{fig:v_ej_micron_vs_Rh}.
The ranges are relatively broad because of the inherent difficulty of determining a single value for the turn back distance
in a continuous brightness distribution.

NEOWISE obtained infrared observations of C/2013~A1 on Jan 16, 2014 \citep{Mainzer_in_prep}, when the comet was at 3.82~AU from the Sun. We used these observations to model the dust production.
Adopting the techniques in \cite{2011ApJ...738..171B}
yields a dust production rate of approximately 10~kg/s 
at the grain velocities for that heliocentric distance, 
and for grains with radius between 1.7~$\mu$m and 2.3~$\mu$m 
\citep{Mainzer_in_prep}.

In this paper we use the HST and NEOWISE observations to anchor the grain ejection velocity and production rate.
In the next Section we discuss how we use these 
constraints to model the velocity of larger dust grains ejected at larger heliocentric distances,
which are in the relevant range for delivery of dust grains to Mars.
In Section~\ref{SEC:RESULTS},
we present the results from numerical simulations where we 
model the dynamical evolution of dust grains after ejection from the comet nucleus
to their close approach with Mars.
In Section~\ref{SEC:DISCUSSION} we discuss the consequences of our modeling choices
and how they likely affect the outcomes of the simulations,
and compare our results with earlier works which characterize
the delivery of dust grains to Mars \citep{2014Icar..231...13M,2014MNRAS.439.3294V,2014arXiv1403.7128Y}.
Conclusion are then drawn in Section~\ref{SEC:CONCLUSIONS}.

\section{Methods}
\label{sec:methods}

The trajectories of the comet nucleus and of the dust grains are
determined numerically by integrating the equations of motion, 
using an efficient numerical integrator with variable timestep \citep{1974CeMec..10...35E},
and that includes gravitational perturbations and solar radiation pressure \citep{1979Icar...40....1B}.
The state vectors of the Sun and the planets are obtained from the DE405 planetary ephemeris \citep{Standish98}.
The initial conditions of the comet nucleus are chosen at the time of the c/a,
as retrieved from the Horizons JPL system \citep{1996DPS....28.2504G},
and are then integrated backward in time to the epoch when the comet was at 10~AU,
i.e., the heliocentric distance that we assume as the beginning of the activity of the comet.
The maximum heliocentric distance of 10~AU is chosen based
on pre-discovery images of the active comet at approximately 8~AU,
and considering that the region of the sky
where the comet was between 8 and 10~AU was not covered by surveys (T.~Spahr, MPC, \emph{priv.~comm.}).
The comet nucleus is considered massless, 
and grains are assigned a radius and density in order to compute the effects of radiation pressure,
and are integrated individually.
As a trade-off between simulation performance and accuracy, 
gravitational interactions are computed only with the Sun, Jupiter, and Mars,
and limit the nominal accuracy of the numerical integrator.
This causes a significant speed-up of simulations compared to including all planets, and causes a difference in c/a distance of dust grains of the order of a few hundred kilometers.
This is relatively small compared to the few thousand kilometers uncertainty on the comet c/a distance,
especially considering the statistical nature of this study.

Dust grains are ejected isotropically from the nucleus,
at a velocity that is determined using an empirical scaling law,
based on the results in \cite{1951ApJ...113..464W}.
This choice is due to the difficulty of developing a detailed self-consistent physical model of grain ejection
driven by hyper-volatiles at large heliocentric distances and then water within approximately 3~AU from the Sun,
as such a model would require some mechanisms for the depletion of hyper-volatiles at shorter heliocentric distance,
or would otherwise produce unrealistically large ejection velocities.
The ejection velocity is:
\begin{equation}
	\frac{v}{v_\text{ref}} = 
	\left[ \frac{\beta}{\beta_{\text{ref}}} \right]^{1/2} \left[ \frac{D}{D_{\text{ref}}} \right]^{-1} =
	\left[ \frac{r}{r_{\text{ref}}} \frac{\rho}{\rho_{\text{ref}}} \right]^{-1/2} \left[ \frac{D}{D_{\text{ref}}} \right]^{-1}
\label{eq:v_ej}
\end{equation}
where we choose the reference grain to have radius $r_{\text{ref}}=1$~mm 
and density $\rho_{\text{ref}} = 1$~g~cm$^{-3}$,
which correspond to $\beta_\text{ref} \simeq 5.76 \times 10^{-4}$,
and $D_{\text{ref}} = 5$~AU for the heliocentric distance.
The value of $D_{\text{ref}}$ is chosen to be approximately halfway between the comet activation and the comet close encounter with Mars.
Since these reference values are fixed for all simulations,
we have $v_\text{ref}$ as the only free parameter characterizing 
the ejection velocity of all grains within each simulation.
Because of the lack of an estimate of the nucleus size,
we consider the ejection velocities defined outside of the gravitational influence region of the nucleus.
In Figure~\ref{fig:v_ej_micron_vs_Rh} we show how the velocities from our modeling
compare to the observed HST velocities.
The three HST observations \citep{Li_in_prep}
indicate velocities of micron-size grains roughly compatible with $v_\text{ref}$ between $0.8$ and $1.0$~m/s.
Because of the significant dispersion in the velocity range,
and the limited number of HST observations, we choose not to fit the exponent 
of the heliocentric distance $D$ in Eq.~\eqref{eq:v_ej} and instead 
scale all velocities as $D^{-1}$ and use different values of $v_\text{ref}$ 
to sample different points of the ejection velocity distribution.
Values of $v_\text{ref}$ above the HST range are reasonable
if we include a high-velocity tail of the distribution due for example to jets.
As such, high $v_\text{ref}$ values would be representative only of a few-percent of the total grains.

The density of dust grains ejected from comets depends strongly on
what they are made up of and the bulk porosity of such dust grains.
In the literature we find estimates as low as approximately 0.1~g/cm$^3$ \citep{1999SSRv...90..149G},
and as high as approximately 2~g/cm$^3$ for inter-planetary dust particles \citep{1994Icar..111..227L},
although that may not necessarily be representative of cometary dust.
In our simulations we use a grain density of 1.0~g/cm$^3$ and provide scaling laws to
different densities for all the main results.

The size-frequency distribution (SFD) and production rate $Q$ of grains are assigned
during post-processing of the simulations.
First, we used a uniform distribution (in log scale) in grain radius between 1~$\mu$m to 10~cm
during the simulations, and a constant production rate.
Then in post-processing each simulated grain is scaled up to represent a far larger number of real grains,
according to the specific SFD and $Q$ distributions selected.
The SFD is parameterized using the differential distribution 
$\text{d}Q / \text{d}r \propto r^{-a}$
and $Q$ depends on the heliocentric distance as
$Q \propto D^{-2}$.
The index $a$ for cometary dust grains depends on many factors
including the heliocentric distance, the specific comet, and the specific 
grain sizes under consideration \citep[e.g.,][and references therein]{2004come.book..555H,2004come.book..565F}. 
However, many authors find that adopting a single power law index 
is convenient for modeling purposes although there are a number of counter 
examples against this approach (e.g., comet 1P/Halley, comet 2P/Encke). 
Therefore, as a compromise, we adopt the values of 3.5 and 4 for $a$, 
which are representative of many comets based on dust tail modeling 
\cite[][and references therein]{2004come.book..565F}.

The flux of grains at Mars is determined from simulations after accounting for SFD and $Q$,
by tracking the grains which come within 3,400~km from the center of Mars.
For scaling purposes, the grains are binned logarithmically into 3 intervals per decade in grain radius,
and in linear intervals of size 0.2~AU in heliocentric distance.
The properties of the grains are assumed to be uniform within each bin.

\section{Results}
\label{SEC:RESULTS}

\begin{figure}[t]
\begin{center}
\includegraphics*[width=\textwidth]{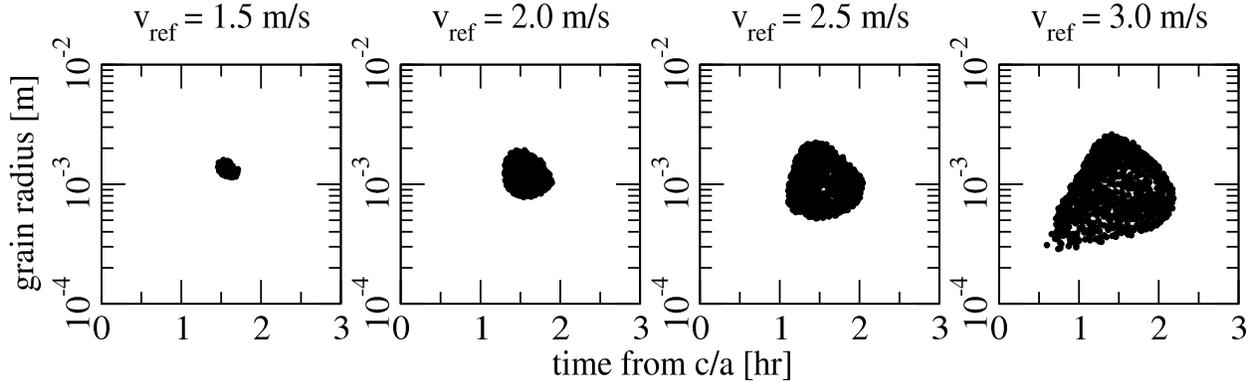}
\end{center}
\caption{Distribution of grain radius and arrival time (in hours) relative to the nucleus close approach time.
The displayed values are for grains with density $\rho_{\text{ref}} = 1$~g~cm$^{-3}$.
For grains with different density $\rho$, the grain radius distribution will shift vertically by a factor 
$\rho_{\text{ref}}/\rho$. 
}
\label{fig:rg_vs_dt_grid}
\end{figure}

We performed several simulations,
each one with a different $v_\text{ref}$ value.
In each simulation, we generated five million grains and tracked their trajectory
as described in Section~\ref{sec:methods}.
Typically only a few hundred to a thousand simulated grains cross the Mars volume, which 
we approximate with a sphere of radius 3,400~km.
In preliminary simulations we used a grain radius between $10^{-6}$ and $10^{-1}$ meter,
and then reduced this range to $10^{-4}$ to $10^{-2}$ meter,
as this is the relevant range for Mars collisions, see Figure~\ref{fig:rg_vs_dt_grid}.
Note how the increased velocity of grains causes a wider range in grain radius to reach Mars,
and also how the arrival time window grows accordingly.
At the highest velocity tested of $v_\text{ref}=3.0$~m/s we start to see smaller grains arriving at early times,
and this effect becomes dominant at higher velocities.
The grains have to be ejected at large heliocentric distances to reach Mars:
beyond 9.1~AU for $v_\text{ref}=1.5$~m/s,
beyond 6.7~AU for $v_\text{ref}=2.0$~m/s,
beyond 4.6~AU for $v_\text{ref}=2.5$~m/s,
and
beyond 2.4~AU for $v_\text{ref}=3.0$~m/s.
Therefore, if the comet was not active at heliocentric distance beyond 8~AU
and the maximum $v_\text{ref}$ was $1.5$~m/s,
no grains could reach Mars.

\begin{figure}[t]
\begin{center}
\includegraphics*[width=\textwidth]{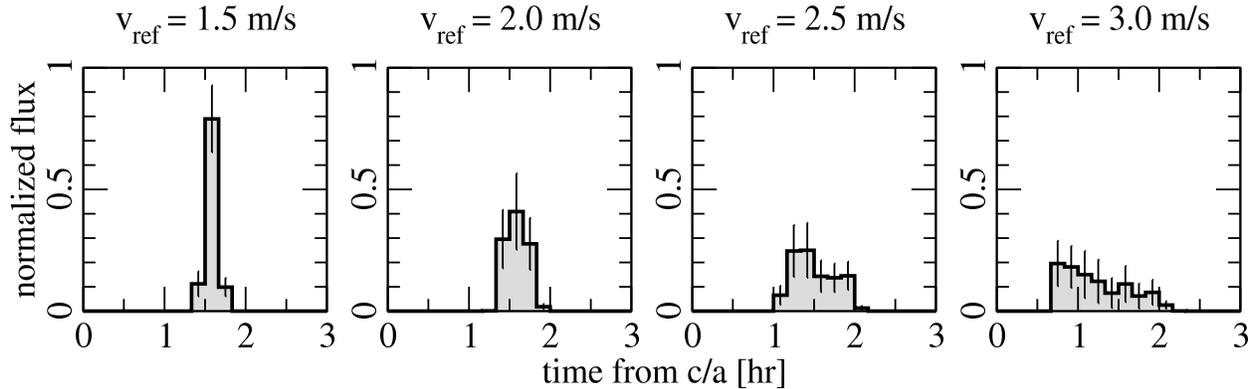}
\end{center}
\caption{Grain count flux normalized to the total fluence and binned over ten-minutes periods,
to represent the fraction of grains colliding with Mars.
The error bars are representative of the counting statistics.
}
\label{fig:flux_histo_vs_dt_grid}
\end{figure}

\begin{figure}[t]
\begin{center}
\includegraphics*[width=0.9\textwidth]{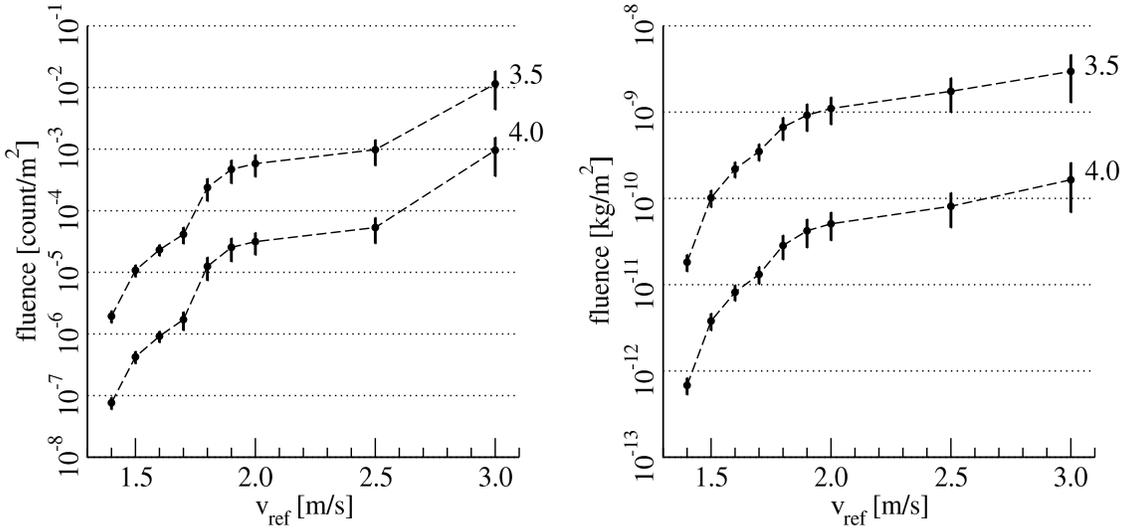}
\end{center}
\caption{Count fluence (left) and mass fluence (right) of models as a function of $v_\text{ref}$,
for SFD index $a=3.5$ and $a=4.0$ as indicated next to each curve.
At velocities lower than displayed, no grains intercept Mars.
Note how the mass fluence flattens at high $v_\text{ref}$ while the count fluence keeps growing,
indicating that the count fluence increase is due to progressively smaller grains.
The displayed values are for grains with density $\rho_{\text{ref}} = 1$~g~cm$^{-3}$,
and for grains with different density $\rho$ the 
count fluence scales as $(\rho_{\text{ref}}/\rho)^{1-a}$,
while the mass fluence scales as $(\rho_{\text{ref}}/\rho)^{3-a}$.
}
\label{fig:fluence_all}
\end{figure}

\begin{figure}[t]
\begin{center}
\includegraphics*[clip=true,width=.9\textwidth]{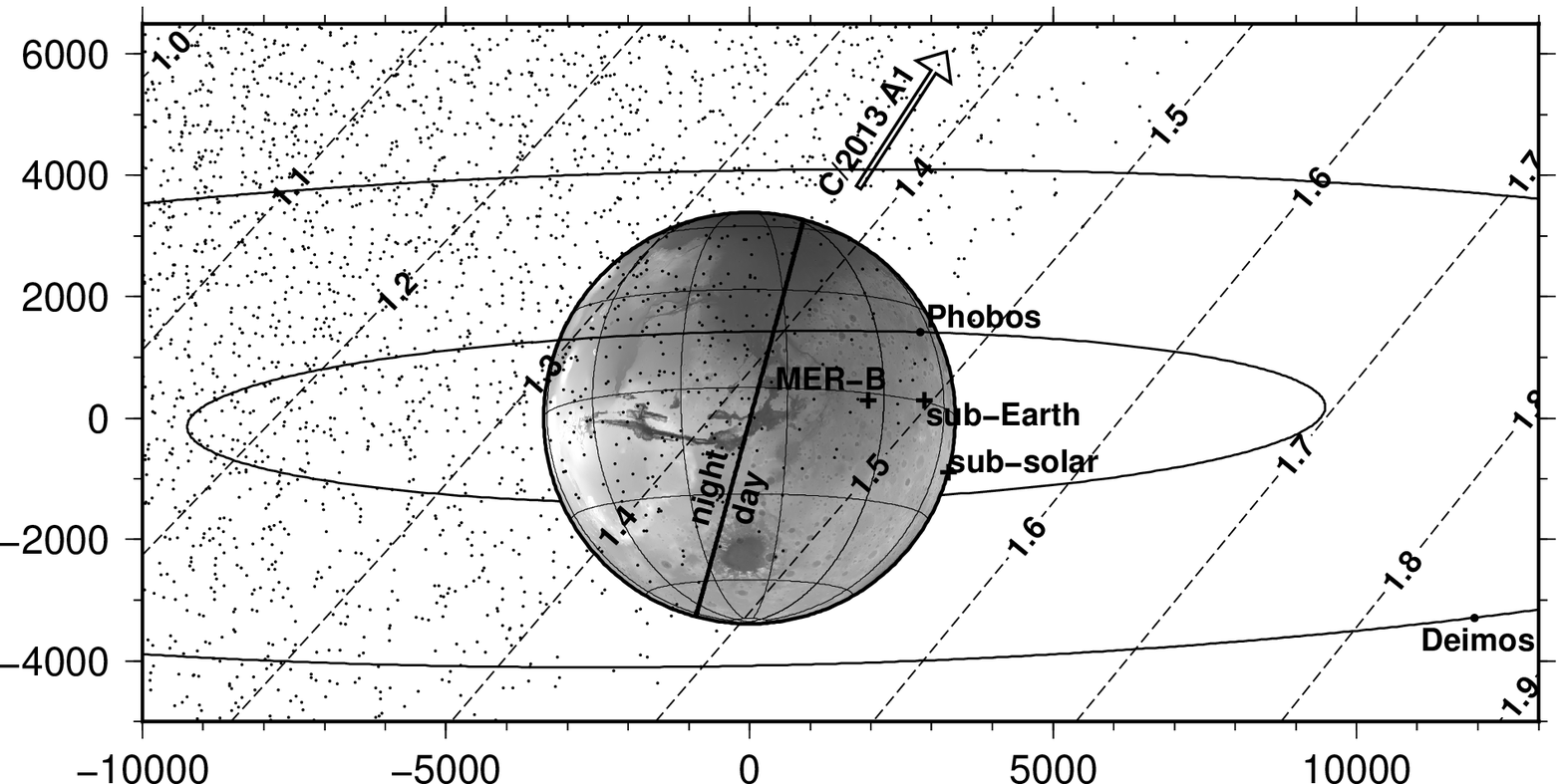}
\end{center}
\caption{View of Mars (north up) centered on latitude $-8.5^\circ$ and longitude $319.4^\circ$ east,
in proximity of Valles Marineris,
which is the position of the sub-radiant of the comet grains flux.
The epoch is October 19, 2014 at 20:04 UT, which is 95 minutes after nominal c/a time, the center of the ten-minutes interval with the peak flux at low $v_\text{ref}$ values.
The oblique dashed curves represent the envelope of the dust grains at the $v_\text{ref}$ value as indicated on each curve, in m/s,
with a sample of $v_\text{ref} = 1.5$~m/s grains included (dots).
The axes indicate the distance from the Mars center in km.
The sub-solar point is just two degrees beyond the edge of Mars.
The Opportunity rover (MER-B) will be in favorable position for possible observations but in daylight.}
\label{fig:mars}
\end{figure}

The arrival time of grains is well concentrated for lower ejection velocities,
with the peak of the flux between 90 and 100 minutes after the close approach time of the nucleus,
see Figure~\ref{fig:flux_histo_vs_dt_grid}.
As we increase the grain ejection velocity,
the arrival time interval tends to widen,
while the peak tends to move to earlier times.
At $v_\text{ref}=3.0$~m/s the peak is well below one hour after c/a, 
and the flux tends to have a fast raise and then a slow decay, over one hour long.

In Figure~\ref{fig:fluence_all} 
we have the integrated fluence (count and mass) versus the grain ejection velocity parameter $v_\text{ref}$
and at different SFD index values.
The figure provides the fluence values for \emph{all} ejected grains at the given $v_\text{ref}$ value.
However, as we have discussed above, HST observations indicate lower velocities at which no grains
intercept Mars.
The way to interpret these plots is then to assume that while the $v_\text{ref}$
parameter from HST indicates the velocity of the bulk of the grains, it is possible 
that the velocity distribution of grains contains an high-velocity tail, 
so that only a few-percent of the grains will have these higher velocities. 
Processes responsible for this could be the presence of jets, or 
the shape of the nucleus causing non-uniformities in the gas flow 
and in the acceleration of grains from the surface.
In view of this interpretation, the fluence values in Figure~\ref{fig:fluence_all} should be
decreased by several orders of magnitude as we move to higher values of $v_\text{ref}$,
to reflect the smaller and smaller fraction of grains which could be in the high-velocity tail.
As a result, the fluence decreases by approximately two orders of magnitude 
at $v_\text{ref} \simeq 1.5$~m/s and by several orders of magnitude at higher $v_\text{ref}$ values,
to obtain a maximum fluence of the order of $10^{-7}$ grains/m$^2$.

The radiant of the dust grains is at R.A.~$41.1^\circ$ and Dec.~$-15.5^\circ$ in the constellation Cetus.
In Figure~\ref{fig:mars} we show the hemisphere of Mars that is exposed to dust grains.
Depending on the velocity of the grains, one side of Mars is accessible at lower velocities than the other one.
Thus, it is reasonable to expect that the flux of grains in the low-$v_\text{ref}$ side
will be somewhat higher than on the opposite side.

\section{Discussion}
\label{SEC:DISCUSSION}

The results presented here can be compared with previous works already in the literature.
In \cite{2014Icar..231...13M} the dust grains fluence is determined 
using both analytical and numerical approaches, obtaining a count fluence of 0.15~m$^{-2}$.
In their model, the comet orbit solution was not yet as constrained as ours,
the grain densities used (0.1~g/cm$^3$) are relatively low,
and the ejection velocities appear significantly higher than what we used.
We find that the modeling details and parameters in \cite{2014Icar..231...13M}
are not supported by observations such as we used in our work,
and the large difference in fluence is a direct consequence of very different models.

In \cite{2014MNRAS.439.3294V},
two different numerical approaches are used to obtain a count fluence of 3.5~m$^{-2}$,
with grains reaching Mars over a five hours period centered at 20~UT.
The model uses very high ejection velocities, 
assumes that water-driven production will be principally responsible for grains delivery to Mars,
and uses very large volumes to compute the fluence,
averaging low flux region and high density regions near the comet nucleus, 
strongly biasing the results towards high fluence values.

In \cite{2014arXiv1403.7128Y}
a dynamical model very similar to ours is employed, 
including a similar velocity scaling and similar constraints from observations,
and an orbit solution just a few weeks older than ours.
They obtain no collisions with Mars at the nominal grain velocities,
and fluence similar to ours if the grain velocity is increased,
after accounting for different parameters (density and SFD index).

Although the peak timing tends to shorter delays after c/a at high $v_\text{ref}$ values (see Figure~\ref{fig:flux_histo_vs_dt_grid}),
the strong attenuation from the velocity distribution causes the $v_\text{ref} \simeq 1.5$~m/s
timing to dominate, with a peak between 90 and 100 minutes after c/a.
The timings provided in this paper refer to the current estimate of the c/a time as indicated in the Introduction, 
and will not change even if the nucleus c/a will change due to possible strong non-gravitational forces which
may become important in the last few months before c/a \citep{Farnocchia_in_prep}.
As a matter of fact, \cite{Farnocchia_in_prep} show that non-gravitational perturbations may become relevant only when approaching the perihelion, while the dust grains that may reach Mars are ejected at larger heliocentric distances.

We did not include fragmentation of grains in this study.
The most likely consequence of fragmentation would be to decrease the probability to reach Mars,
because of the increased radiation pressure effects.

\section{Conclusions}
\label{SEC:CONCLUSIONS}

We have presented the results of our dynamical model for the delivery of dust grains 
ejected from the nucleus of comet C/2013~A1 (Siding Spring) to Mars.
Direct extrapolation of the grain ejection velocity from HST observations indicates that the bulk of the grains will likely miss Mars.
If we include the possibility of a few-percent of the grains to have higher velocities,
we find that millimeter radius grains ejected at 9~AU or more from the Sun will collide with Mars.
At higher velocities, younger grains from sub-millimeter to several millimeter can reach Mars too.
The maximum fluence will be of the order of $10^{-7}$ grains/m$^2$.
The timing of the peak flux is expected to be 90--100 minutes after c/a,
which is between 19:59 and 20:09 UT of October 19, 2014.

\acknowledgments

This research was supported by a contract to the Planetary Science Institute 
by the NASA JPL Mars Critical Data Products Program
and by NASA Planetary Atmospheres Program grant NNX11AD91G to the Planetary Science Institute.
HST data and funding were provided through GO-13610.
NEOWISE is a project of JPL/Caltech funded by the Planetary Science Division of NASA.
The work of D. Farnocchia was conducted at JPL/Caltech under a contract with NASA.
R.~Stevenson is funded by the NASA Postdoctoral Program.

\end{document}